\documentclass{article}
\usepackage{spconf,amsmath,graphicx,hyperref}
\usepackage{amssymb}  
\usepackage{siunitx}
\usepackage{bbm}  
\usepackage{multirow}
\usepackage{tabularray}

\title{PRETRAINED CONFORMERS FOR AUDIO FINGERPRINTING AND RETRIEVAL}

\name{Kemal Altwlkany\textsuperscript{1,2}, Elmedin Selmanovic\textsuperscript{2}, Sead Delalic\textsuperscript{2}\thanks{\textit{https://github.com/KemalAltwlkany/pretrained-conformers}}}
\address{\textsuperscript{1}Infobip\\\textsuperscript{2}Faculty of Science, University of Sarajevo}

\begin{document}
\ninept

\maketitle

\begin{abstract}
Conformers have shown great results in speech processing due to their ability to capture both local and global interactions. In this work, we utilize a self-supervised contrastive learning framework to train conformer-based encoders that are capable of generating unique embeddings for small segments of audio, which generalize well to previously unseen data. We obtain results comparable to existing approaches for audio retrieval tasks, while using only 3 seconds of audio to generate embeddings. Our models are almost completely immune to temporal misalignments and retain good performance in cases of audio distortions such as noise, reverb or extreme temporal stretching. Code and models are made publicly available and the results are easy to reproduce as we train and test using popular and freely available datasets of different sizes.
\end{abstract}


\section{Introduction}

Given a small excerpt of an audio file, the task of content-based audio retrieval is to identify the original file from which the excerpt was taken. Content-based audio retrieval is often performed using audio fingerprinting techniques. That is, for every audio file of interest, a unique fingerprint is generated by taking smaller segments of the file and generating low-dimensional representations of the segments. These low-dimensional representations collectively make up the audio fingerprint. Retrieval using an excerpt is performed by computing the low-dimensional representations of the excerpt and finding corresponding matches, often facilitated through vector search. Retrieval is limited to audio files for which fingerprints have been made and added to a database, while extending the database is performed by simply updating it with new fingerprints. Note that these low-dimensional representations do not necessarily need to be embeddings (or more generally vectors), instead, they can be as simple as time-frequency pairs within the spectrogram, as is usually the case in more classical audio fingerprinting approaches \cite{wang2003industrial}.

The most common application of audio fingerprinting is in music identification services, such as those offered by Apple (Shazam) \cite{wang2003industrial}, Google (Now Playing \cite{gfeller2017now}), SoundHound \cite{soundhoundonline} and others. Audio fingerprinting is also used for detecting copyright infringements or tracking advertisements \cite{wang2003industrial,chang2021neural}. In telecommunications, audio fingerprinting has been employed to search for duplicate audio when identifying unsolicited phone calls \cite{prasad2020s,ellis2011echoprint} or even to improve network performance by quickly identifying announcements played during session initiation \cite{altwlkany2024knowledge,pribic2025systems}.

\begin{figure}[t]
  \centering
  \includegraphics[width=\linewidth]{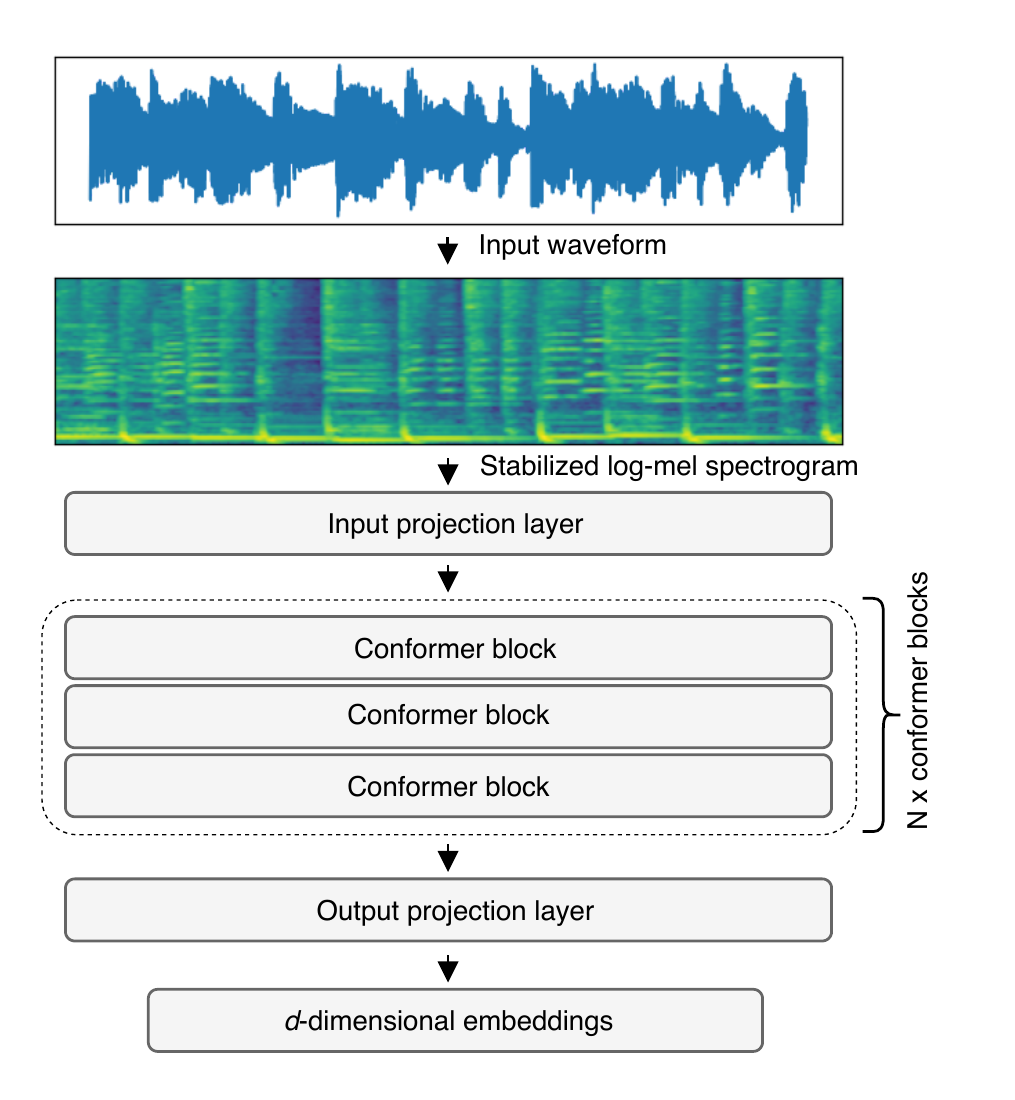}
  \caption{Encoder architecture. We compute the stabilized log-mel spectrogram and pass it through a linear projection layer, followed by several ($N$) conformer blocks, before a final projection layer that outputs $d$-dimensional embeddings.}
  \label{fig:model-architecture}
\end{figure}

In \cite{cortes2025peaknetfp} the authors classify audio fingerprinting techniques into three categories: local descriptors-based \cite{haitsma2002highly,covell2007known,drevo2014dejavu}, peak-based \cite{wang2003industrial,joren2014panako,sonnleitner2015robust} and neural-based approaches \cite{gfeller2017now,chang2021neural,singh2023simultaneously}.
Neural-based approaches are mainly based on contrastive learning frameworks and convolutional neural networks (CNNs) \cite{gfeller2017now,chang2021neural}. Recent advances such as PeakNetFP \cite{cortes2025peaknetfp} combine peak extraction techniques with contrastive learning. GraFPrint \cite{bhattacharjee2025grafprint} is the first approach to utilize Graph Neural Networks (GNNs), while the authors of \cite{singh2022attention} demonstrate an attention-based approach.

Our contributions can be summarized as follows; we propose the first conformer-based architecture for neural audio fingerprinting, achieving competitive results for audio retrieval tasks. CNNs exploit local features effectively, while transformers are good at capturing content-based global interactions and have achieved state-of-the-art results in speech recognition \cite{gulati20interspeech,islam2024comprehensive}. Conformers combine the best of both worlds, which makes them a suitable choice for audio fingerprinting tasks; when recognizing audio, the spectral content itself is just as important as its positioning in time within the audio. After an initial pretraining phase, our models learn to generalize well and can generate unique embeddings for previously unseen data. We highlight the importance of data augmentation during training, especially the task of generating difficult contrastive examples to prevent the models from overfitting. The paper is accompanied by three models of different sizes which we open source together with the code which enables easy replication of our results.

\section{Method}

Figure \ref{fig:model-architecture} provides a high-level overview of our proposed conformer-based architecture.

\subsection{Conformer encoder}
We present three models, which are limited by the number of their parameters: small (1.5M), medium (8.8M) and large (26.2M). The model architectures, spectral parameters and hyperparameters are summarized in Table \ref{tab:conformer-hparams}. All of our encoders share the high-level architecture from Figure \ref{fig:model-architecture}; they compute the stabilized log-mel spectrogram from an input audio waveform and propagate it through the network. The output is a $d$-dimensional embedding, which is $\ell_{2}$ normalized. All models were trained using the Adam optimizer \cite{kingma2014adam} (default values; $\beta_{1}=0.9$, $\beta_{2}=0.999$, $\epsilon=1e-8$) and a dropout rate $P_{drop}=0.5$. The learning rate was kept constant for the small and medium models, while a simple hybrid adaptive strategy was applied that gradually decreased the learning rate while increasing the batch size during training of the large model.

\subsection{Contrastive learning}
\begin{figure}[t]
  \centering
  \includegraphics[width=\linewidth]{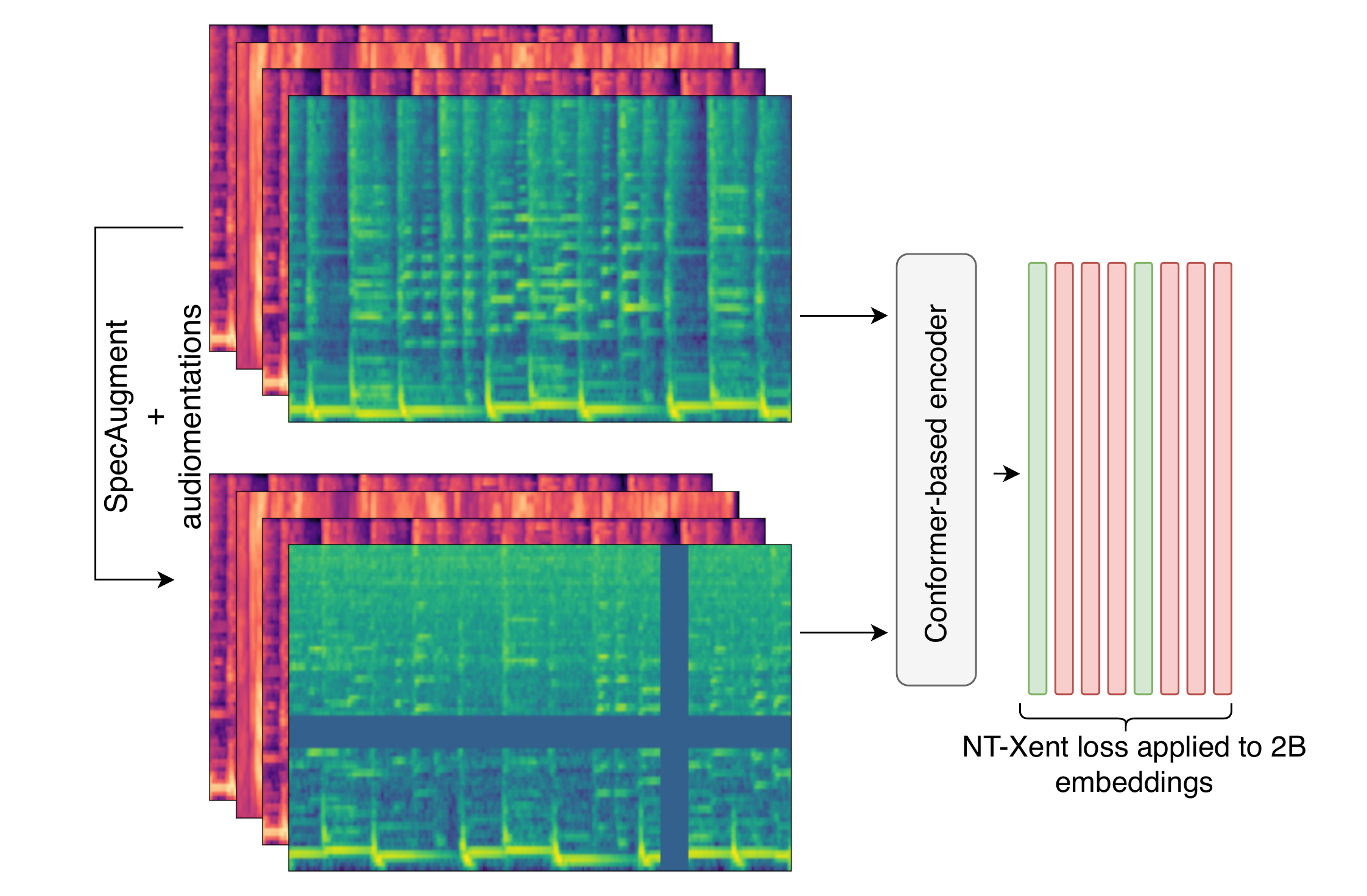}
  \caption{Training process. For a $B$-sized mini-batch, we augment each sample and pass the original and augmented versions through our encoder resulting in a total of $2B$ embeddings. We compute the \textit{NT-Xent} loss by treating the original and its derived augmented replica as positive pairs, whereas the remaining $2(B-1)$ samples are treated as negatives. The bottom green spectrogram is an augmented version of the top one, it is temporally shifted to the left, it's spectral content is pitch shifted (upward), it contains blue noise and some frequencies and time points have been masked.}
  \label{fig:training-process}
\end{figure}

Similar to previous work \cite{gfeller2017now,chang2021neural,cortes2025peaknetfp,bhattacharjee2025grafprint}, we use the SimCLR self-supervised contrastive learning framework \cite{chen2020simple} in which we wish to minimize the normalized temperature-scaled cross entropy loss (NT-Xent). The training process is illustrated in Figure \ref{fig:training-process}. Given a mini-batch of $B$ audio segments (samples), we generate a replica for each of the $B$ samples by applying various audio augmentation techniques. The original ${z_{i}}$ and replica $z_{j}$ are considered positive pairs, while the remaining $2(B-1)$ samples are all treated as negative examples. In literature, $\mathrm{sim}(z_{i}, z_{j}) = z_{i}^\top z_{j} / \Vert z_{i} \Vert \Vert z_{j}\Vert$ is commonly used to denote the cosine similarity between two samples $z_{i}$ and $z_{j}$, which is equal to the dot product in our case, as we $\ell_{2}$ normalize the embeddings at the output of our model. The loss for a positive pair of samples $(i, j)$ is computed as:
\begin{equation}
\ell_{i,j} = -\log \frac{\exp\left( \mathrm{sim}({z_i}, z_j) / \tau \right)}{\sum_{k=1}^{2B} \mathbbm{1}_{[k \ne i]} \exp\left( \mathrm{sim}({z_i}, {z_j}) / \tau \right)}
\label{eq:nt_xent}
\end{equation}

\noindent where $\mathbbm{1}_{[k \ne i]} \in \{0, 1\}$ evaluates to $1$ iff $k \neq i$ and $\tau$ is the temperature parameter. The loss is now computed for each positive pair, both $(i, j)$ and $(j, i)$, and averaged into the total cost function.

\begin{table}[b]
    \tiny
    \centering
    \caption{Model architecture, spectral and hyper-parameters.}
    \resizebox{0.9\linewidth}{!}{%
\centering
\begin{tabular}{l|rrr}
Model                           & \multicolumn{1}{l}{Small} & \multicolumn{1}{l}{Medium} & \multicolumn{1}{l}{Large}  \\ 
\hline
Num Params (M)                  & 1.5                         & 8.8                          & 26.2                          \\
Encoder Dim                     & 256                         & 512                          & 768                          \\
Encoder Layers                  & 2                         & 3                          & 4                          \\
Attention heads                 & 4                        & 8                         & 12                         \\
Conv kernel size                & 5                        & 5                         & 5                         \\ 
Embeddings dim                & 128                         & 128                          & 128                          \\
\hline
Sample rate                 & 16k                        & 16k                         & 16k                         \\
Input len (sec)                 & 3                        & 3                         & 3                         \\
FFT len                         & 1024                        & 1024                         & 1024                         \\
Hop len                         & 128                        & 128                        & 128                         \\
N mels                          & 80                         & 80                          & 96                          \\ 
\hline
Training set (samples)                          & 6.4k                         & 20k                          & 83k         \\
Epochs                          & 100                         & 100                          & 100                          \\
NT-Xent $\tau$ & 0.07                        & 0.05                         & 0.05                         \\
Learning rate                   & 1e-4                         & 2e-4                          & variable                          \\
Batch size                      & 64                         & 64                          & variable                          \\
Weight decay ($\lambda$)                    & 5e-4                         & 5e-4                          & 5e-4                          
\end{tabular}
    }
    \label{tab:conformer-hparams}
\end{table}

\subsection{Data}

Our models are trained using the Free Music Archive (FMA) \cite{defferrard2016fma} which provides predefined data subsets of different sizes. We use the small subset (8k tracks) to train the small model (1.5M), medium (25k tracks) to train the medium model (8.8M), and large (106k tracks) for the large model (26.2M). Each of the subsets has an already predefined train-test-validation split that we follow.

\subsection{Generating positive pairs}
As shown in Figure \ref{fig:training-process}, the self-supervised training process consists of selecting $B$ samples from the training set and applying data augmentation techniques to generate replicas, whose distance in the embedding space we wish to minimize from the distance of the originals.

We rely on the Python audiomentations library \cite{iver2025zenodo} to augment data; adding background noise at various SNR rates (datasets ESC-50 \cite{piczak2015esc} and MS-SNSD \cite{reddy19_interspeech}), adding colored noise at various SNR and decay rates, applying impulse response/reverb (dataset MIT IR Survey \cite{traer2016statistics}), pitch-shifting the audio, applying polarity inversion, applying $\tanh$ distortion, time-stretching the audio and time-shifting the audio. Despite using various audio augmentations, we observed that SpecAugment \cite{park19e_interspeech} was crucial to prevent overfitting, especially for our large model.

\subsubsection{Beta-distributed temporal shifting}

The input to all models is a 3-second long audio segment, which means that during training we randomly select such a segment from each audio taken into the batch. For a sample rate of \SI{16}{\kilo\hertz} this amounts to a segment of \SI{48}{k} samples.

In practice, it is desirable to allow for slight temporal mismatches when performing audio retrieval, i.e., the query segment does not have to be temporally aligned with the retrieved segment. Initially, we opted for a maximum of 5\% mismatch (\SI{150}{\milli\second} or \SI{2400} samples), thus we would randomly shift the replica (augmented audio) during training between 0 and 2400 samples. Although no overfitting occurred, that is, no substantial deviations between training and test/validation recall, the initial retrieval rates of models trained under such circumstances for relatively large time shifts, e.g. \SI{120}{\milli\second}, were not satisfying. Simply put, our models were underfit for tasks with large time shifts.

\begin{figure}[t]
  \centering
  \includegraphics[width=\linewidth]{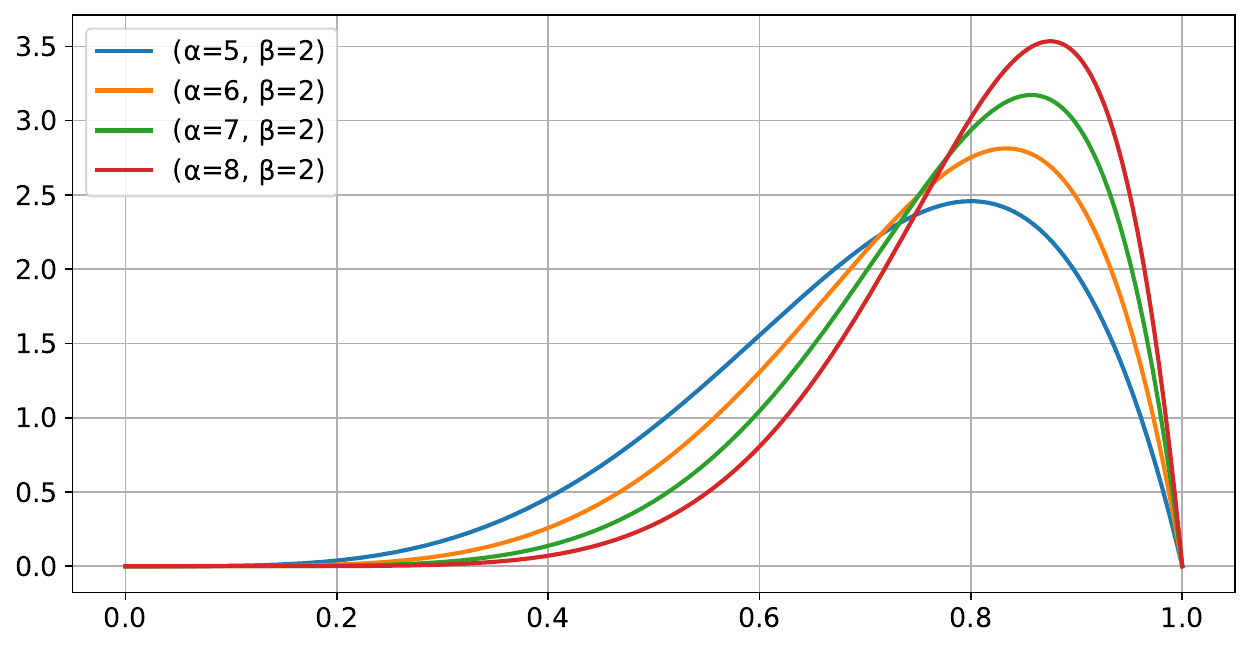}
  \caption{Probability density function of the beta distribution for values $\beta = 2$ and $\alpha = [5, 6, 7, 8]$. Given $\alpha=8$ and a maximum value of \SI{150}{\milli\second}, the mean would amount to \SI{120}{\milli\second}.}
  \label{fig:beta-pdf}
\end{figure}

After some investigation, similar to FaceNet \cite{schroff2015facenet}, we concluded that it is important to train the model on what the authors refer to as \textit{hard} examples. Thus, the training process was adjusted to encourage larger temporal shifts (harder examples) by sampling the time offset by which we shift audio from a beta distribution instead of a uniform distribution. As illustrated in Figure \ref{fig:beta-pdf}, sampling from a beta distribution with parameters $\alpha=8$ and $\beta=2$ with the minimum and maximum offset values of \SI{0}{\milli\second} and \SI{150}{\milli\second} will on average apply an offset of \SI{120}{\milli\second}. This seemingly small change was crucial to improving performance and making our models almost completely robust to any temporal misalignments ranging from \SI{0}{\milli\second} to \SI{150}{\milli\second}.

\section{Experiments}

After training each model on its respective training dataset, we compute the embeddings for the large test dataset of FMA \cite{defferrard2016fma}. Faiss \cite{johnson2019billion} is used to store and retrieve the embeddings. Results regarding retrieval time or analyses regarding different indexing mechanisms of Faiss are left out since: (1) our work mainly focused on developing a conformer-based model for generating audio fingerprints, not an end-to-end audio retrieval application, and (2), Faiss is almost a standard for retrieval of audio embeddings, so similar results can be expected as in other approaches which utilize Faiss \cite{chang2021neural,cortes2025peaknetfp,bhattacharjee2025grafprint}.

All reported results represent an average over 5 identical test runs. In a single test, we perform 5 queries for every audio file by selecting 5 linearly spaced 3-second long excerpts of each audio. As is commonly done in audio fingerprinting literature, we report on the top-1 and top-5 hit rate, often referred to as the \textit{exact} and \textit{near} matches or \textit{accuracy} \cite{chang2021neural,bhattacharjee2025grafprint,singh2022attention}. For easier distinction, the top-5 hit rate will further be enlisted inside parentheses.

\begin{figure}[t]
  \centering
  \includegraphics[width=\linewidth]{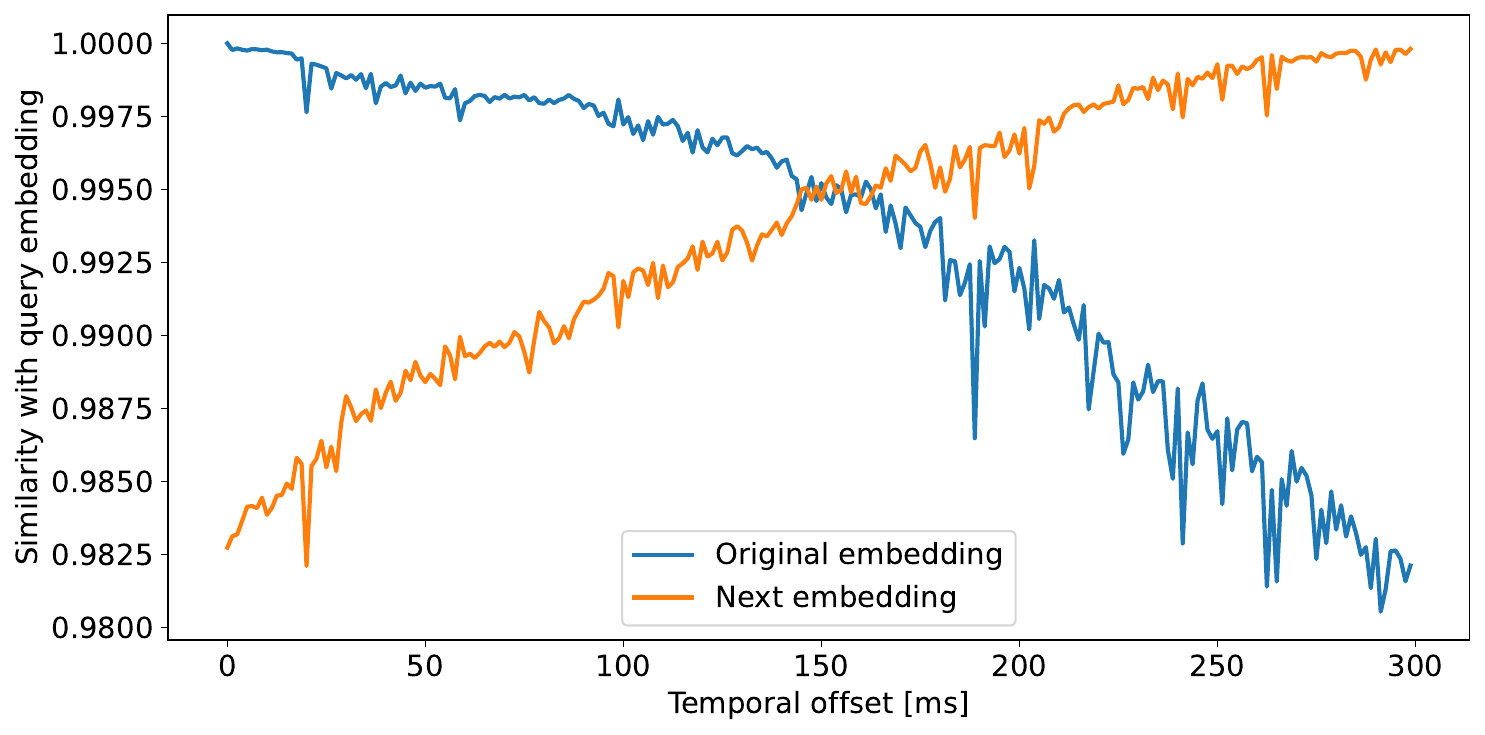}
  \caption{Robustness to temporal shifts. For a given query embedding, we plot the similarity between itself (original) and the next embedding, which are spaced \SI{300}{\milli\second} apart. We gradually increase the temporal shift of the query (which now acts as a replica) and observe how the similarity between the replica and next embedding increases, while the similarity between the replica and original decreases. Note how the similarity curves intersect at around \SI{150}{\milli\second}.}
  \label{fig:distance-between-embeddings}
\end{figure}

\subsection{Performance of conformer models}
Of all distortions, our models are most robust to temporal shifts, which is supported by maintaining near-perfect results in the presence of such distortions, especially in terms of the top-5 hit rate. In Table \ref{tab:temporal-distortions}, we provide the hit rates achieved when applying temporal shifts, expressed as a percentage of half the distance from the next fingerprint (i.e. \SI{300}{\milli\second}). Our models perform well in this type of distortion as they have learned to continuously decrease the similarity metric between two audio segments as the time distance between them increases, mimicking a monotonic function. Figure \ref{fig:distance-between-embeddings} illustrates this desirable property of the embedding's robustness to temporal misalignments.

In Table \ref{tab:temporal-distortions} we further provide the hit rates when dealing with reverb. Although these results are good, especially for the large conformer, the models are not as robust to reverb as to temporal offsets, which is further supported by Table \ref{tab:noise-distortions} in which the overall performance for background and colored noises at various SNR levels is provided. Note that the results were evaluated using ESC-50 and the test split of MS-SNSD \cite{piczak2015esc,reddy19_interspeech}, which none of the models had seen during training.

\begin{table}[h]
    \centering
    \caption{Top-1 (top-5) hit rates for time-distorted queries.}
    \resizebox{\linewidth}{!}{%
\centering
\begin{tabular}{l|llll|l}
       & \multicolumn{4}{c|}{Temporal shift (\% of \SI{300}{\milli\second})}                                                         & \multicolumn{1}{c}{\multirow{2}{*}{Reverb}}  \\
Model  & \multicolumn{1}{c}{10\%} & \multicolumn{1}{c}{20\%} & \multicolumn{1}{c}{30\%} & \multicolumn{1}{c|}{40\%} & \multicolumn{1}{c}{}                         \\ 
\hline
Small  & 98.7 (100)               & 98.8 (100)               & 98.5 (100)               & 98.4 (100)                & 88.8 (91.4)                                  \\ 
\hline
Medium & 98.6 (100)               & 98.6 (100)               & 98.4 (100)               & 98.3 (100)                & 92.2 (94.8)                                  \\ 
\hline
Large  & 98.6 (100)               & 98.7 (100)               & 98.4 (100)               & 98.2 (100)                & 96.2 (98.2)                                 
\end{tabular}
}
    \label{tab:temporal-distortions}
\end{table}

\begin{table*}
    \centering
    \caption{Comparison of our large conformer model (26.2M) with other audio fingerprinting approaches: top-1 hit rate.}
    \resizebox{\linewidth}{!}{%
\centering

\begin{tabular}{l|r|r|r|r|rrr|r|l}
\multicolumn{1}{l}{}     & \multicolumn{1}{l}{}              & \multicolumn{1}{l}{}                 & \multicolumn{1}{l}{}            & \multicolumn{1}{l}{}           & \multicolumn{4}{c}{Top-1 hit rate (SNR)}                                                &                                                               \\ 
\hline
Model                    & \multicolumn{1}{l|}{Query length} & \multicolumn{1}{l|}{Embeddings size} & \multicolumn{1}{l|}{Batch size} & \multicolumn{1}{l|}{FMA subset} & \multicolumn{1}{l}{\SI{5}{\dB}} & \multicolumn{1}{l}{\SI{10}{\dB}} & \multicolumn{1}{l|}{\SI{15}{\dB}} & \multicolumn{1}{l|}{No noise} & Comment                                                       \\ 
\hline
Conformer (26.2M)        & 3                               & 128                                  & arbitrary                              & \textbf{large}                           & \textbf{93.9}                    & \textbf{97.1}                     & \textbf{97.8}                      & \textbf{98.8}                          & Up to \SI{120}{\milli\second} temporal shift                                 \\
\hline
NAFP \cite{chang2021neural} & 5                                 & 128                                  & 640                             & large                           & /                       & $\geqslant92$                        & $\geqslant92$                         & $\geqslant92$                            & Varying SNR (\SI{0}{}-\SI{10}{\dB}) and reverb                                                           \\
Now-playing \cite{gfeller2017now,chang2021neural}             & 5                                 & 64                                   & unknown                               & large                           & /                       & $\geqslant 79.8$                        & $\geqslant 79.8$                         & $\geqslant 79.8$                          & Implemented by NAFP \cite{chang2021neural} under same conditions                              \\
Dejavu \cite{chang2021neural,drevo2014dejavu}                  & 6                                 & /                                    & /                              & small                           & /                       & /                        & /                         & 69.6                          & Reported by NAFP \cite{chang2021neural}                                               \\ 
\hline
Audfprint \cite{singh2022attention,ellis20142014}                & 3                                 & /                                    & /                              & medium                          & 88.5                    & 91.4                     & 93.8                      & /                             & Implemented by Attention-based \cite{singh2022attention}                  \\
Audfprint \cite{singh2022attention,ellis20142014}               & 5                                 & /                                    & /                              & medium                          & 89.4                    & 91.5                     & 94.2                      & /                             & Implemented by Attention-based \cite{singh2022attention}                  \\
NAFP \cite{chang2021neural,singh2022attention}          & 3                                 & 128                                  & 512                             & medium                          & 88.5                    & 91.4                     & 93.8                      & /                             & Implemented by Attention-based \cite{singh2022attention}                                      \\
\hline
Attention-based \cite{singh2022attention}          & 3                                 & 128                                  & 512                             & medium                          & 92.9                    & 94.6                     & 95.5                      & /      & /                                                            \\
Attention-based \cite{singh2022attention}         & 5                                 & 128                                  & 512                             & medium                          & 93.9                    & 94.7                     & 96.8                      & /   & /                                                                \\ 
FE+HT \cite{singh2023simultaneously} & 3 & 128 (16-quantized) & / & medium & 90.8 & 95.5 & 96.3 & / & Up to \SI{50}{\milli\second} temporal shift\\
\hline
GraFPrint \cite{bhattacharjee2025grafprint}               & 1                                 & 128                                  & 256                             & large                           & 61.8                    & 71.6                     & 83.8                      & /     & /                          \\
GraFPrint \cite{bhattacharjee2025grafprint}                & 3                                 & 128                                  & 256                             & \textbf{medium}                          & \textbf{98.3}                    & \textbf{99.3}                     & \textbf{99.6}                      & /           & /                                    \\
Attention-based \cite{bhattacharjee2025grafprint,singh2022attention}         & 3                                 & 128                                  & unknown                         & medium                          & 88.4                    & 92.6                     & 95.2                      & /                             & Implemented and reported by GraFPrint \cite{bhattacharjee2025grafprint}                        
\end{tabular}
    }
    \label{tab:comparison-to-others}
\end{table*}

\begin{table}[ht]
    \centering
    \caption{Top-1 (top-5) hit rates for noise-distorted queries.}
    \resizebox{\linewidth}{!}{%
\centering
\begin{tabular}{c|ccc|ccc}
\multicolumn{1}{l|}{} & \multicolumn{3}{c|}{Background noise} & \multicolumn{3}{c}{Colored noise}              \\ 
Model  & SNR 5dB     & SNR 10dB    & SNR 15 dB   & SNR 5 dB    & SNR 10dB    & SNR 15dB     \\ 
\hline
Small  & 83.4 (86.2) & 92.9 (95.0) & 96.0 (97.9) & 77.8 (81.0) & 88.7 (91.3) & 93.5 (95.8)  \\ 
\hline
Medium & 87.8 (90.6) & 95.0 (97.1) & 96.8 (98.8) & 91.8 (94.4) & 95.6 (97.8) & 97.0 (99.0)  \\ 
\hline
Large  & 93.9 (96.2) & 97.1 (99.0) & 97.8 (99.6) & 95.6 (97.8) & 97.3 (99.2) & 97.9 (99.7)
\end{tabular}
    }
    \label{tab:noise-distortions}
\end{table}

\begin{figure}[t]
  \centering
  \includegraphics[width=\linewidth]{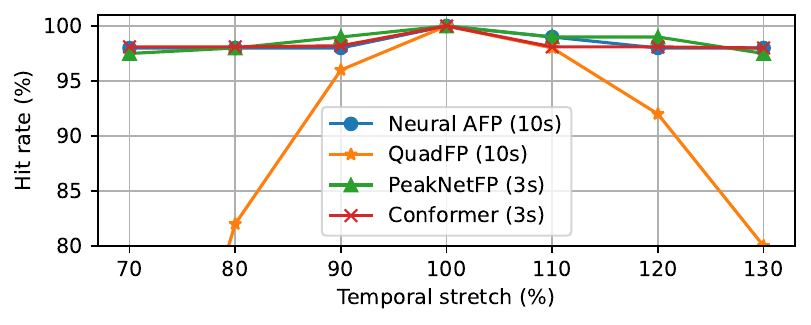}
  \caption{Top-1 hit rate vs. temporal stretch.}
  \label{fig:temporal-stretch}
\end{figure}

\subsection{Comparison with other fingerprinting approaches}

\subsubsection{Temporal distortions}
We compare the performance of our best model with the current state-of-the-art Neural audio fingerprinter (NAFP) \cite{chang2021neural} and PeakNetFP \cite{cortes2025peaknetfp} for time-stretching. These models can handle extreme audio slowdowns or speed-ups and have improved significantly compared to the previous state-of-the-art QuadFP \cite{sonnleitner2015robust}. Our model seems to perform very slightly under the performance of PeakNetFP at 90\% and 110\% time stretching, but it is marginally better at the more extreme rates of 70\% and 130\%, as shown in Figure \ref{fig:temporal-stretch}. Note that QuadFP and NAFP use a query sample that is more than 3 times longer compared to PeakNetFP and the conformer.

\subsubsection{Noise distortions}
In Table \ref{tab:comparison-to-others}, we compare the performance of our best model with the current state-of-the-art approaches in noisy conditions. It is difficult to discern and point out a clear winner, since there are many variables at play, e.g., the length of the query sample used, the size of the embeddings or the subset of FMA \cite{defferrard2016fma}. The audio distortions are also not uniform, in NAFP \cite{chang2021neural} the results were reported under a varying SNR range, whereas in \cite{singh2023simultaneously,bhattacharjee2025grafprint,singh2022attention} independent test runs were repeated by fixing these values. The most important differences are listed under column \textit{Comments} inside Table \ref{tab:comparison-to-others}.

\subsection{Discussion}
For noisy audio, GraFPrint \cite{bhattacharjee2025grafprint} claims to achieve the best results with a given query length of only 3 seconds, evaluated on the medium subset of FMA. NAFP is the most referenced and tested implementation and has been confirmed to achieve similar results by several sources \cite{chang2021neural,cortes2025peaknetfp,bhattacharjee2025grafprint,singh2022attention}. Aside being evaluated on the large subset of FMA, another advantage of NAFP is that it achieves state-of-the-art performance in cases of extreme time stretching, while GraFPrint does not report on the results for such distortions.

Given the results from Table \ref{tab:comparison-to-others} and Figure \ref{fig:temporal-stretch}, we conclude that our large conformer model (26.2M) is comparable to the current state-of-the-art in terms of extreme temporal distortions (NAFP \cite{chang2021neural}, PeakNetFP \cite{cortes2025peaknetfp}) using a small query length of 3 seconds. Regarding noisy distortions, the large model has comparable performance to GraFPrint \cite{bhattacharjee2025grafprint}, especially at SNR rates of \SI{10}{\dB} and \SI{15}{\dB}, while being the best performing model evaluated on the large subset of FMA \cite{defferrard2016fma}. Thus, our model achieves good results on both frontiers.

From Table \ref{tab:temporal-distortions} and Table \ref{tab:noise-distortions}, we conclude that the small (1.5M) and medium (8.8M) models do not exhibit state-of-the-art performance, except for temporal shifts. However, given their reduced size and complexity, these models are very suitable for applications in which no severe audio distortions are expected, being marginally outperformed by the large model in such circumstances.

\section{Conclusion}
Conformers excel at capturing both local and global interactions, making them a natural candidate for audio fingerprinting techniques. We showed that pretrained conformer encoders are capable of generating distinct embeddings for small segments of previously unseen audio which remain robust even in cases of severe audio distortions. Our best performing model matches the current state-of-the-art, while our smaller models achieve good results under lighter distortions. In future work, we aim to investigate quantization techniques, both on the encoder weights, as well as the embeddings generated by the encoder and to provide a framework for easier comparison and verification of audio fingerprinting techniques.

\vspace*{2mm}

\bibliographystyle{IEEEbib}
\bibliography{refs}

\end{document}